# Hamilton-Jacobi Fractional Sequential Mechanics

Eqab M. RABEI[*] and Bashar S. ABABNEH
Department of Physics, Mutah University, Al-Karak, Jordan

## Abstract

As a continuation of Rabei et al. work [11], the Hamilton- Jacobi partial differential equation is generalized to be applicable for systems containing fractional derivatives. The Hamilton- Jacobi function in configuration space is obtained in a similar manner to the usual mechanics. Two problems are considered to demonstrate the application of the formalism. The result found to be in exact agreement with Agrawal's formalism.

*eqabrabei@yahoo.com

# 1-Introduction

The Hamiltonian formulation of non-conservative systems has been developed by Riewe[1,2].He used the fractional derivative [3,4,5] to construct the Lagrangian and Hamiltonian for non-conservative systems. As a sequel to Riewe's work, Rabei et al. [6] used Laplace transforms of fractional integrals and fractional derivatives to develop a general formula for the potential of any arbitrary forces, conservative or non-conservative. This led directly to the consideration of the dissipative effects in Lagrangian and Hamiltonian formulations. Besides, the canonical quantization of non-conservative systems carried out by Rabei et al. [7].

Other investigations and further developments are given by Agrawal [8] .He presented the fractional variational problems and the resulting equations are found to be similar to those for variation problems containing integral order derivatives. This approach is extended to classical fields with fractional derivatives [9]. Besides, Kilmek [10] showed that the fractional Hamiltonian is usually not a constant of motion, even in the case when the Hamiltonian is not an explicit function of time. In addition, as a continuation of Agrawal's work [8], Rabei et al. [11] achieved the passage from the Lagrangian containing fractional derivatives to the Hamiltonian. The Hamilton's equations of motion are obtained in a similar manner to the usual mechanics.

In the present work, the Hamilton – Jacobi partial differential equation (HJPDE) is generalized to be applicable for systems containing fractional derivatives.

The paper is organized as follows: In Sec. 2 Lagrangian and Hamiltonian formalisms with fractional derivatives are reviewed briefly. In Sec.3, Hamilton-Jacobi Partial differential equations with fractional derivatives is constructed, and two illustrative examples are given in Sec. 4.

## 2- Hamiltonian Formalism with Fractional Derivative

Several definitions of a fractional derivative have been proposed. These definitions include Riemann–Liouville, Grünwald–Letnikov, Weyl, Caputo, Marchaud, and Riesz fractional derivatives. Here; the problem is formulated in terms of the left and the right Riemann–Liouville fractional derivatives.

The left Riemann–Liouville fractional derivative defined as

$$
{}_aD_x^\alpha f(x) = \frac{1}{\Gamma(n-\alpha)} \left(\frac{d}{dx}\right)^n \int_a^x (x-\tau)^{n-\alpha-1} f(\tau) d\tau \tag{1}
$$

Which is denoted as the LRLFD and the right Riemann–Liouville fractional derivative reads as

$$
{}_xD_b^\alpha f(x) = \frac{1}{\Gamma(n-\alpha)} \left(-\frac{d}{dx}\right)^n \int_x^b (x-\tau)^{n-\alpha-1} f(\tau) d\tau \tag{2}
$$

Which is denoted as the RRLFD. Here $\alpha$ is the order of the derivative such that $n-1 \leq \alpha \leq n$ and $\Gamma$ represents the Euler gamma function. If $\alpha$ is an integer, these derivatives are defined in the usual sense, i.e.

$$
{}_aD_x^\alpha f(x) = \left(\frac{d}{dx}\right)^\alpha f(x) \quad , \quad {}_xD_b^\alpha f(x) = \left(-\frac{d}{dx}\right)^\alpha f(x) \quad , \quad \alpha = 1,2,3,.... \tag{3}
$$

The fractional operator ${}_aD_x^\alpha$ can be written as [13]

$$
{}_aD_x^\alpha = \frac{d^n}{dx^n} {}_aD_x^{\alpha-n} \tag{4}
$$

Where the number of additional differentiations n is equal to [α] +1, where [α] is the whole part of α. The operator ${}_aD_x^\alpha$ is a generalization of differential and integral operators and can be introduced as follows:

$$
{}_aD_x^\alpha = \begin{cases} \dfrac{d^\alpha}{dx^\alpha} & \operatorname{Re}(\alpha) > 0 \\ 1 & \operatorname{Re}(\alpha) = 0 \\ \displaystyle\int_a^x (d\tau)^{-\alpha} & \operatorname{Re}(\alpha) < 0 \end{cases} \tag{5}
$$

Following to Agrawal [8], the Euler-Lgrange equations for fractional calculus of variations problem is obtained as

$$\frac{\partial L}{\partial q} + {}_t D_b^\alpha \frac{\partial L}{\partial {}_a D_t^\alpha q} + {}_a D_t^\beta \frac{\partial L}{\partial {}_t D_b^\beta q} = 0 \tag{6}$$

Where L is the genaralized Lagrangian function of the form $L(q, {}_a D_t^\alpha q, {}_t D_b^\beta q, t)$

The generalized momenta are introduced as

$$p_\alpha = \frac{\partial L}{\partial {}_a D_t^\alpha q} \quad , \quad p_\beta = \frac{\partial L}{\partial {}_t D_b^\beta q} \tag{7}$$

And the Hamiltonian depending on the fractional time derivatives reads as

$$H = p_\alpha {}_a D_t^\alpha q + p_\beta {}_t D_b^\beta q - L \tag{8}$$

In Ref [11], the Hamilton's equations of motion are obtained in a similar manner to the usual mechanics. These equations read as,

$$\frac{\partial H}{\partial t} = -\frac{\partial L}{\partial t} \quad ; \quad \frac{\partial H}{\partial p_\alpha} = {}_a D_t^\alpha q \quad , \quad \frac{\partial H}{\partial p_\beta} = {}_t D_b^\beta q \quad ; \quad \frac{\partial H}{\partial q} = {}_t D_b^\alpha p_\alpha + {}_a D_t^\beta p_\beta$$

It is observed that the fractional Hamiltonian is not a constant of motion even though the Lagrangian does not depend on the time explicitly.

## 3. Hamilton-Jacobi Partial Differential Equation with Fractional Derivatives

In this section, the determination of the Hamilton-Jacobi partial differential equation for systems with fractional derivatives is discussed. According to Rabei et al. [11], the fractional Hamiltonian is written as

$$H\big(q, p_\alpha, p_\beta, t\big) = p_\alpha \,_a D_t^\alpha q + p_\beta \,_t D_b^\beta q - L(q, \,_a D_t^\alpha q, \,_t D_b^\beta q, t) \tag{9}$$

Consider the canonical transformation with a generating function

$$F_2\big(\,_a D_t^{\alpha-1} q, \,_t D_b^{\beta-1} q, P_\alpha, P_\beta, t\big)$$

Then, the new Hamiltonian will take the form

$$K\big(Q, P_\alpha, P_\beta, t\big) = P_\alpha \,_a D_t^\alpha Q + P_\beta \,_t D_b^\beta Q - L'(Q, \,_a D_t^\alpha Q, \,_t D_b^\beta Q, t) \tag{10}$$

The old canonical coordinates $q, p_\alpha, p_\beta$, satisfy the fractional Hamilton's principle that can be put in the form

$$\delta \int_{t_1}^{t_2} \big(p_\alpha \,_a D_t^\alpha q + p_\beta \,_t D_b^\beta q - H\big) dt = 0 \tag{11}$$

At the same time the new canonical coordinates $Q, P_\alpha, P_\beta$, of course satisfy a similar principle.

$$\delta \int_{t_1}^{t_2} \big(P_\alpha \,_a D_t^\alpha Q + P_\beta \,_t D_b^\beta Q - K\big) dt = 0 \tag{12}$$

The simultaneous validity of Eq. (11) and Eq. (12) does not mean of course that the integrands in both expressions are equal. Since the general form of the Hamilton's principle has zero variation at the end points, both statements will be satisfied if the integrands connected by a relation of the form [12]

$$p_\alpha \,_a D_t^\alpha q + p_\beta \,_t D_b^\beta q - H = P_\alpha \,_a D_t^\alpha Q + P_\beta \,_t D_b^\beta Q - K + \frac{dF}{dt} \tag{13}$$

Where the function F is given as:

$$F = F_2\left({}_aD_t^{\alpha-1}q, {}_tD_b^{\beta-1}q, P_\alpha, P_\beta, t\right) - P_\alpha {}_aD_t^{\alpha-1}Q - P_\beta {}_tD_b^{\beta-1}Q \qquad (14)$$

The function $F_2$ is called Hamilton's principal function $S$ for a contact transformation.

$$F_2 = S\left({}_aD_t^{\alpha-1}q, {}_tD_b^{\beta-1}q, P_\alpha, P_\beta, t\right) \qquad (15)$$

Thus,

$$\frac{dF}{dt} = \frac{dS}{dt} - \frac{dP_\alpha}{dt}\,{}_aD_t^{\alpha-1}Q - P_\alpha\frac{d}{dt}\,{}_aD_t^{\alpha-1}Q - \frac{dP_\beta}{dt}\,{}_tD_b^{\beta-1}Q - P_\beta\frac{d}{dt}\,{}_tD_b^{\beta-1}Q$$

By using definitions of fractional calculus given in Eq. (4) then we have

$$\frac{dF}{dt} = \frac{dS}{dt} - \frac{dP_\alpha}{dt}\,{}_aD_t^{\alpha-1}Q - P_\alpha\,{}_aD_t^{\alpha}Q - \frac{dP_\beta}{dt}\,{}_tD_b^{\beta-1}Q - P_\beta\,{}_tD_b^{\beta}Q \qquad (16)$$

Substituting the values of the $\dfrac{dF}{dt}$ from Eq. (16) into the Eq. (13) we have

$$p_\alpha\,{}_aD_t^{\alpha}q + p_\beta\,{}_tD_b^{\beta}q - H = -K + \frac{dS}{dt} - \frac{dP_\alpha}{dt}\,{}_aD_t^{\alpha-1}Q - \frac{dP_\beta}{dt}\,{}_tD_b^{\beta-1}Q \qquad (17)$$

But

$$\frac{dS}{dt} = \frac{\partial S}{\partial\,{}_aD_t^{\alpha-1}q}\frac{d}{dt}\,{}_aD_t^{\alpha-1}q + \frac{\partial S}{\partial\,{}_tD_b^{\beta-1}q}\frac{d}{dt}\,{}_tD_b^{\beta-1}q + \frac{\partial S}{\partial P_\alpha}\frac{dP_\alpha}{dt} + \frac{\partial S}{\partial P_\beta}\frac{dP_\beta}{dt} + \frac{\partial S}{\partial t}$$

Again using definitions of fractional calculus given in Eq. (4) we have the following form

$$\frac{dS}{dt} = \frac{\partial S}{\partial\,{}_aD_t^{\alpha-1}q}\,{}_aD_t^{\alpha}q + \frac{\partial S}{\partial\,{}_tD_b^{\beta-1}q}\,{}_tD_b^{\beta}q + \frac{\partial S}{\partial P_\alpha}\frac{dP_\alpha}{dt} + \frac{\partial S}{\partial P_\beta}\frac{dP_\beta}{dt} + \frac{\partial S}{\partial t} \qquad (18)$$

Substituting the values of the $\dfrac{dS}{dt}$ from Eq. (18) into the Eq. (17) we get

$$p_\alpha = \frac{\partial S}{\partial\,{}_aD_t^{\alpha-1}q} \qquad , \qquad p_\beta = \frac{\partial S}{\partial\,{}_tD_b^{\beta-1}q} \qquad (19)$$

$$_aD_t^{\alpha-1}Q = \frac{\partial S}{\partial P_\alpha} \qquad , \qquad _tD_b^{\beta-1}Q = \frac{\partial S}{\partial P_\beta} \tag{20}$$

$$H + \frac{\partial S}{\partial t} = K \tag{21}$$

We can automatically ensure that the new variables are constant in time by requiring that the transformed Hamiltonian K shall be identically zero, In other words, $Q, P_\alpha, P_\beta$ are constants. We see by putting K = 0 that this generating function must satisfy the partial differential equation.

$$H + \frac{\partial S}{\partial t} = 0 \tag{22}$$

This equation is called the Hamilton –Jacobi equation. Let us assume that

$$P_\alpha = E_1 \qquad , \qquad P_\beta = E_2$$

Where $E_1$, $E_2$ are constants. Then the action function (15), can be expressed as

$$S = S\left(_aD_t^{\alpha-1}q, \, _tD_b^{\beta-1}q, E_1, E_2, t\right) \tag{23}$$

Further insight into the physical significance of Hamilton's principal function S is furnished by an examination of the total time derivative, which can be computed from the formula

$$\frac{dS}{dt} = \frac{\partial S}{\partial \, _aD_t^{\alpha-1}q} \, _aD_t^\alpha q + \frac{\partial S}{\partial \, _tD_b^{\beta-1}q} \, _tD_b^\beta q + \frac{\partial S}{\partial t} \tag{24}$$

By using Eq. (19) we have

$$\frac{dS}{dt} = p_\alpha \, _aD_t^\alpha q + p_\beta \, _tD_b^{\beta-1}q - H$$

And using Eq. (9) we have

$$\frac{dS}{dt} = L$$

Thus

$$S = \int_{t_1}^{t_2} L \, dt \qquad (25)$$

If we restrict our considerations to the time -independent Hamiltonians, then the Hamilton-Jacobi function can be written in the form

$$S = W_1\left(_a D_t^{\alpha-1} q, E_1\right) + W_2\left(_t D_b^{\beta-1} q, E_2\right) + f\left(E_1, E_2, t\right) \qquad (26)$$

Where $W$ is called Hamilton's characteristic function and the function, $f$ takes the following form:

$$f\left(E_1, E_2, t\right) = -Et$$

Making use of equations (19) and (20) we obtain:

$$p_\alpha = \frac{\partial W_1}{\partial \, _a D_t^{\alpha-1} q} \qquad , \qquad p_\beta = \frac{\partial W_2}{\partial \, _t D_b^{\beta-1} q} \qquad (27)$$

$$_a D_t^{\alpha-1} Q = \frac{\partial W_1}{\partial E_1} = \lambda_1 \qquad , \qquad _t D_b^{\beta-1} Q = \frac{\partial W_2}{\partial E_2} = \lambda_2 \qquad (28)$$

Here $\lambda_1$ , $\lambda_2$ are constants.

The physical significance of W can be understood by writing its total time derivative

$$\frac{dW_1}{dt} = \frac{\partial W_1}{\partial \, _a D_t^{\alpha-1} q} \, _a D_t^{\alpha} q \qquad (29)$$

Comparing this expression to the results of substituting Eq. (27) into Eq. (29) we see that

$$\frac{dW_1}{dt} = p_\alpha \, _a D_t^{\alpha} q \qquad \Rightarrow \qquad W_1 = \int p_\alpha \, _a D_t^{\alpha} q \, dt \qquad \Rightarrow \qquad W_1 = \int p_\alpha d \, _a D_t^{\alpha-1} q \qquad (30)$$

Again one may show that

$$W_2 = \int p_\beta \, d \, _t D_b^{\beta-1} q \qquad (31)$$

# 4. Illustrative Examples

To demonstrate the application of our formalism, let us discuss the following models:

As a first model consider the lagrangian given by Agrawal [8]

$$L = \frac{1}{2} \left( {}_0D_t^{\alpha} q \right)^2$$

The (HJPDE) for this Lagrangian is calculated as

$$\frac{1}{2} \left( p_\alpha \right)^2 + \frac{\partial S}{\partial t} = 0$$

Using Eq. (27) we obtain

$$\frac{1}{2} \left( \frac{\partial W_1}{\partial {}_0D_t^{\alpha-1} q} \right)^2 - E = 0$$

Solving this equation we have

$$W_1 = \sqrt{2E} \; {}_0D_t^{\alpha-1} q$$

Thus

$$p_\alpha = \sqrt{2E}$$

Making use of Eq. (26) we obtain the function S as:

$$S = \sqrt{2E} \; {}_0D_t^{\alpha-1} q - Et$$

Eq. (28) leads to

$$_0D_t^{\alpha-1} Q = \frac{\partial S}{\partial E} = \frac{1}{\sqrt{2E}} \; {}_0D_t^{\alpha-1} q - t = \lambda_1$$

Thus

$$_0D_t^{\alpha-1} q = \sqrt{2E} \left( t + \lambda_1 \right)$$

Or

$$_0D_t^{\alpha} q = \sqrt{2E} = p_\alpha$$

This is the same result obtained by Rabei et al. [11], which is equivalent to Agrawal formalism [8].

As a second model consider the Lagrangian given by Rabei et al. [11]

$$L = \frac{1}{2} \left( {}_0 D_t^{\alpha} q \right)^2 + \frac{1}{2} \left( {}_t D_1^{\beta} q \right)^2 + {}_0 D_t^{\alpha} q \, {}_t D_1^{\beta} q$$

The Hamiltonian is calculated as

$$H = \frac{1}{2} \left( p_{\alpha} \right)^2 = \frac{1}{2} \left( p_{\beta} \right)^2$$

Thus, the Hamilton-Jacobi partial differential equation reads as:

$$\frac{1}{2} \left( p_{\alpha} \right)^2 + \frac{\partial S}{\partial t} = 0$$

Making use of Eq. (26) we have

$$\frac{1}{2} \left( \frac{\partial W_1}{\partial {}_0 D_t^{\alpha-1} q} \right)^2 - E = 0$$

Thus,

$$W_1 = \sqrt{2E} \; {}_0 D_t^{\alpha-1} q$$

Again the (HJPDE) can be written as

$$\frac{1}{2} \left( p_{\beta} \right)^2 + \frac{\partial S}{\partial t} = 0$$

Then

$$\frac{1}{2} \left( \frac{\partial W_2}{\partial {}_t D_1^{\beta-1} q} \right)^2 - E = 0$$

Which leads to

$$W_2 = \sqrt{2E} \; {}_t D_1^{\beta-1} q$$

Thus the Hamilton-Jacobi action function can be written as

$$S = \sqrt{2E} \; {}_0 D_t^{\alpha-1} q + \sqrt{2E} \; {}_t D_1^{\beta-1} q - Et$$

Where

$$p_{\alpha} = \frac{\partial W_1}{\partial {}_0 D_t^{\alpha-1} q} = \sqrt{2E}$$

And

$$p_{\beta} = \frac{\partial W_2}{\partial {}_t D_1^{\beta-1} q} = \sqrt{2E}$$

Using Eq. (28) we get

$${}_0 D_t^{\alpha-1} Q = \frac{1}{\sqrt{2E}} \; {}_0 D_t^{\alpha-1} q + \frac{1}{\sqrt{2E}} \; {}_t D_1^{\beta-1} q - t = \lambda_1$$

Thus

$${}_0 D_t^{\alpha-1} q + {}_t D_1^{\beta-1} q = \sqrt{2E} \left( t + \lambda_1 \right)$$

Or

$$_0D_t^\alpha q +_t D_1^\beta q = \sqrt{2E}$$

Then

$$p_\alpha =_0 D_t^\alpha q +_t D_1^\beta q$$

And

$$p_\beta =_0 D_t^\alpha q +_t D_1^\beta q$$

These Leads to

$$(_0D_t^\beta +_t D_1^\alpha)(_0 D_t^\alpha q +_t D_1^\beta q) = 0$$

This result is in exact agreement with Rabei et al. [11].

## 5- Conclusion

In This work we have studied the Hamilton-Jacobi partial differential equation for systems containing fractional derivatives. A general theory to solve the Hamilton-Jacobi partial differential equation is proposed for systems containing fractional derivatives under the condition that this equation is separable. The Hamilton-Jacobi function is determined in the same manner as for usual systems. Finding this function enables us to get the solutions of the equations of motion.

In order to test our formalism, and to get a somewhat deeper understanding, we have examined two examples of systems with fractional derivatives. The result found to be in exact agreement with Lagrangian formulation given by Agrawal [8] and with Hamiltonian formulation given by Rabei et al. [11].

## 6- References